\documentclass[submission, Proceedings]{SciPost}

\hypersetup{
    colorlinks,
    linkcolor={red!50!black},
    citecolor={blue!50!black},
    urlcolor={blue!50!black}
}

\usepackage[bitstream-charter]{mathdesign}
\DeclareSymbolFont{usualmathcal}{OMS}{cmsy}{m}{n}
\DeclareSymbolFontAlphabet{\mathcal}{usualmathcal}

\usepackage{lineno}
\linenumbers

\begin{document}

\nolinenumbers

\ \\[-3cm] {\begin{flushright} JLAB-THY-21-3465 \end{flushright}}

\begin{center}{\Large \textbf{
CJ15 global PDF analysis with new electroweak data \\
from the STAR and SeaQuest experiments
}}\end{center}

\begin{center}
Sanghwa Park\textsuperscript{1$\star$},
Alberto Accardi\textsuperscript{2,3},
Xiaoxian Jing\textsuperscript{3,4},
J.F. Owens\textsuperscript{5}
\end{center}

\begin{center}
{\bf 1} Stony Brook University, Stony Brook, NY, USA
\\
{\bf 2} Hampton University, Hampton, VA, USA
\\
{\bf 3} Jefferson Lab, Newport News, VA, USA
\\
{\bf 4} Southern Methodist University, Dallas, TX, USA
\\
{\bf 5} Florida State University, Tallahassee, FL, USA
\\


* sanghwa.park@stonybrook.edu
\end{center}

\begin{center}
\today
\end{center}


\definecolor{palegray}{gray}{0.95}
\begin{center}
\colorbox{palegray}{
  \begin{tabular}{rr}
  \begin{minipage}{0.1\textwidth}
    \includegraphics[width=22mm]{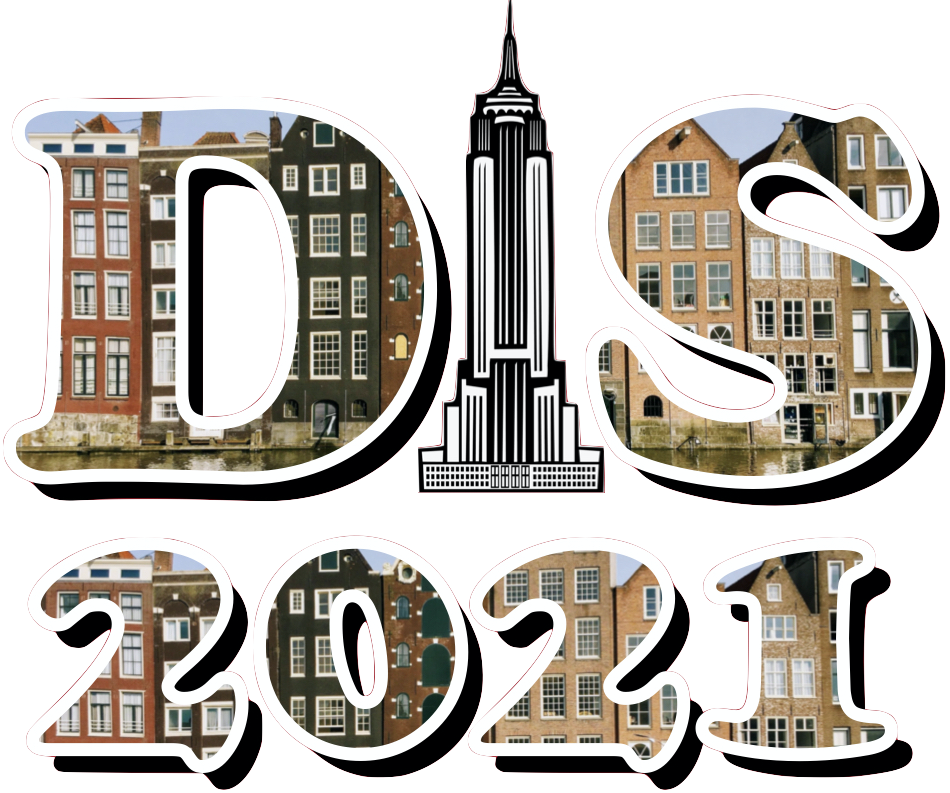}
  \end{minipage}
  &
  \begin{minipage}{0.75\textwidth}
    \begin{center}
    {\it Proceedings for the XXVIII International Workshop\\ on Deep-Inelastic Scattering and
Related Subjects,}\\
    {\it Stony Brook University, New York, USA, 12-16 April 2021} \\
    \doi{10.21468/SciPostPhysProc.?}\\
    \end{center}
  \end{minipage}
\end{tabular}
}
\end{center}

\thispagestyle{empty}

\section*{Abstract}
{\bf
We present updates to a recent CTEQ-Jefferson Lab (CJ) global analysis of parton distribution functions with a new set of electroweak data that provide unique access to quark flavor separation in the proton. In particular, recent $\boldsymbol W$ and $\boldsymbol Z$ boson measurements from the STAR experiment at RHIC put additional constraints on light quarks and antiquarks near the valence regime. The new measurement of the Drell-Yan lepton pair production ratio 
in $\boldsymbol{p+p}$ and $\boldsymbol{p+d}$ collisions 
by the SeaQuest experiment at Fermilab extends the large-$\boldsymbol x$ coverage of the previous E866 experiment and sheds new light on the light antiquarks distribution. In this report, the impact of these new data sets on parton distribution functions will be presented with emphasis given to the flavor asymmetry of the light antiquark sea at large values of the parton momentum $\boldsymbol x$.
}





\section{Introduction}
\label{sec:intro}

The flavor asymmetry of the light antiquark sea inside the proton is of great importance to understand the structure of the nucleon, see, for example, \cite{GARVEY2001203,Geesaman:2018ixo} for a review. 
In particular, an excess of $\bar{d}$ over $\bar{u}$ antiquarks has been observed at moderate values of the parton momentum fraction $x$ in Deep Inelastic Scattering and Drell-Yan experiments on proton and deuteron targets. In particular, the measurement of the Drell-Yan production ratio in $p+p$ and $p+d$ collisions by E866 \cite{PhysRevLett.80.3715,PhysRevD.64.052002}
suggested a potential sign change of the flavor asymmetry ($\bar{d}-\bar{u}$) at $x \approx 0.25$ region, albeit with large uncertainty -- a behavior that is difficult to understand theoretically and has generated a lot of discussion in the past 20 years. 

New data released this year will afford us a fresh look into this issue. 
Recent measurements by the SeaQuest collaboration \cite{Dove:2021ejl} provide new insight into the light antiquark distributions by extending the large $x$ coverage of earlier experiments.
At the same time new data on $W$ boson production in $p+p$ collisions by the STAR collaboration \cite{PhysRevD.103.012001} offer an alternative way to probe the light antiquark sea in the proton at $x \approx 0.2$.
Indeed, due to the $V-A$ structure of the weak interaction the $W$ boson only couples to left-handed quarks and right-handed antiquarks, and correlates the charge of the produced $W$ with initial state quark flavors. The cross sections and charge ratio of $W^+$ and $W^-$ bosons production can thus provide direct sensitivity to the flavor separated quark distribution functions. Unlike the Drell-Yan and DIS measurements which use deuteron targets to extract $\bar{d}/\bar{u}$, the STAR measurement is free from nuclear effects and naturally probes the proton quarks at a larger scale $Q^2 = M_W^2$ set by the $W$ boson mass -- but at slightly smaller values of $x$ than the SeaQuest experiment.

Here we report recent studies of light antiquark sea distributions performed in the CTEQ-JLab (CJ) collaboration's global QCD analysis framework using, for the first time, the new $W$ boson cross section measurements by STAR and Drell-Yan measurements by SeaQuest.

\section{Weak boson production in proton-proton collisions}

At leading order in perturbative QCD (pQCD), the charge ratio of $W$ cross sections is sensitive to light quarks and antiquarks,
\begin{equation}
    \frac{\sigma_{W^{+}}}{\sigma_{W^{-}}} \ \approx \ \frac{u(x_1)\bar{d}(x_2)+\bar{d}(x_1)u(x_2)}{d(x_1)\bar{u}(x_2)+\bar{u}(x_1)d(x_2)}
   \ \underset{y_{_W} \approx 0}{\approx} \ \frac{\bar d / \bar u}{d / u } 
\end{equation}
where $x_{1(2)} = (M_{_W}/\sqrt{s}) \exp(\pm y_{_W})$ is the fractional momentum carried by the scattering partons, with $y_{_W}$ the rapidity of the produced boson. At midrapidity, where $x_1=x_2=M_{_W}/\sqrt s$, the cross section ratio directly accesses the antiquark and quark ratios.

At the Relativistic Heavy Ion Collider (RHIC), where  measurements have been performed at $\sqrt{s}$ = 500 and 510 GeV \cite{PhysRevD.103.012001}, $W$ bosons are identified via their lepton $W^\pm\rightarrow e^\pm\nu$ decays, without detecting the accompanying (anti)neutrino. Combined with decay kinematics, the observed $W$ cross sections at lepton pseudorapidity $-1 \leq \eta_e \leq 1.5$ are sensitive to the unpolarized quark distributions in the region $0.05 < x < 0.3$. We will also discuss $Z$ boson production, for which the boson rapidity can be fully reconstructed via the dilepton $Z \rightarrow e^+e^-$ decay at $|y_{_{Z}}| \leq 1$.

\section{Drell-Yan lepton pair production in proton-hadron collisions}
At leading order, the Drell-Yan lepton pair production in proton-hadron collisions can be described as the annihilation of a quark-antiquark pair into a virtual photon which then decays into a pair of leptons. Assuming charge symmetry and small deuteron nuclear corrections, the cross section ratio for lepton pair production in $p+p$ and $p+d$ collisions in the forward region ($x_F = x_b - x_t \gg 0$) can be approximated by
\begin{equation}
    \frac{\sigma^{pd}}{\sigma^{pp}}\approx 1 + \frac{\bar{d}(x_t)}{\bar{u}(x_t)}
\end{equation}
where $x_b$ and $x_t$ are the fractions of the dilepton pair momentum relative to, respectively, the momentum of the beam and of the target hadrons. 

The E866 and SeaQuest experiments both used a proton beam against LD$_2$ and LH$_2$ targets. E866 measured the ratio in the kinematic range of $0.015 < x_t < 0.3$ and an average $Q^2$ of 54 GeV$^2$. The recently released SeaQuest data covers the range of $0.1 < x_t < 0.45$ and $Q^2$ of 22--40 GeV$^2$, with improved precision and an evaluation of correlated systematic uncertainties.

\section{Light Antiquarks in the CJ15 Global Analysis}

The latest CJ15 global fit (CJ15) has been performed using the world DIS data set, as well as a variety of jet and electroweak boson production measurements \cite{Accardi:2016qay}. Among these, Drell-Yan measurements by the Fermilab's E866 experiment provide the strongest constraints on the light antiquark sea, covering the $0.015 
\lesssim x < \lesssim 0.3$ parton momentum region.

At the input scale of $Q^2_0 = 3$ GeV$^2$, a standard five-parameter functional form is used for most of parton species, including the $\bar u + \bar d$ combination: 
\begin{equation}
    xf(x, Q^2_0) = a_0 x^{a_1} (1-x)^{a_2}(1+a_3 \sqrt{x} + a_4 x) \ .
\label{eq:CJ15-par}
\end{equation}
In the original \texttt{CJ15} fit, due to the limited $x$ coverage of the E866 data and the sharp downturn these required of the $\bar{d}/\bar{u}$ ratio, this was directly parametrized as 
\begin{equation}
    \bar{d}/\bar{u} = a_0 x^{a_1}(1-x)^{a_2} + 1 + a_3 x(1-x)^{a_4} \ ,
\end{equation}
enforcing the theoretical expectation from most modeling efforts that $\bar d/\bar u \to 1$ as $x \to 1$. This assumption was however revisited in Ref.~\cite{Accardi:2019ofk}, where the $\bar d - \bar u$ difference was considered instead of $\bar d / \bar u$ and parametrized as in Eq.~\eqref{eq:CJ15-par}, and the resulting fit was called \texttt{CJ15-a}.
Even if the new parametrization allowed for it, no strong indication of a sign change in the $\bar d - \bar u$ asymmetry in the $x \lesssim 0.3$ region measured by E866 was, however, found.

With the new STAR data sensitive to the smaller-$x$ rise of the $\bar d - \bar u$ asymmetry, and the new SeaQuest data constraining the $\bar d / \bar u$ ratio at $0.15 \lesssim x \lesssim 0.4$, well across the region where E866 indicated  this would drop below 1, we  can now revisit this whole issue. We therefore performed a new fit tentatively called \texttt{CJ15-a+} that utilized the more flexible parametrization of \texttt{CJ15-a} but included the STAR $W^+/W^-$ cross section ratio and the SeaQuest Drell-Yan proton to deuteron target cross section ratio on top of the standard CJ15 data set. (The impact of the charge separated $W$ cross section and of the $Z$ cross section on the determination of light quark PDF was found to be minimal, and these data excluded from the fit.) The results are discussed in the next Section.

\section{Results}

\begin{figure}
\centering
\includegraphics[width=0.49\textwidth]{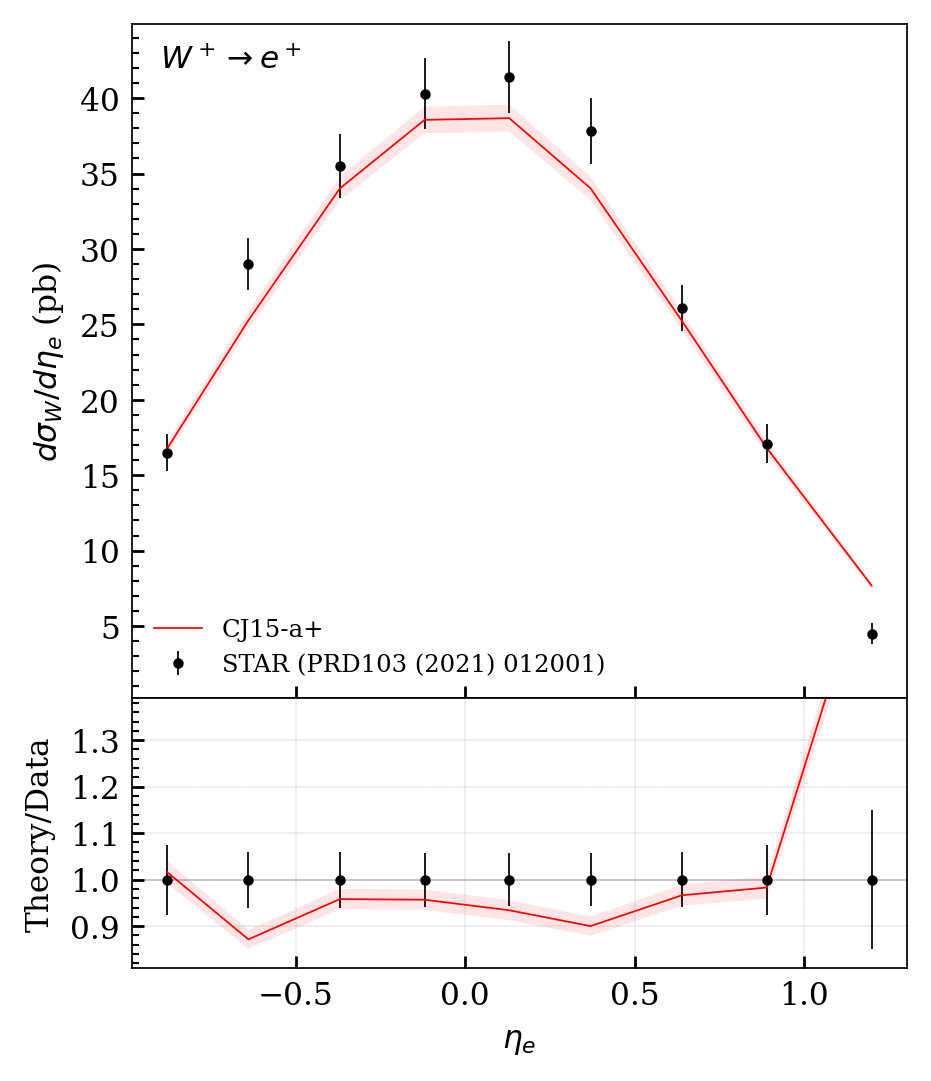}
\includegraphics[width=0.49\textwidth]{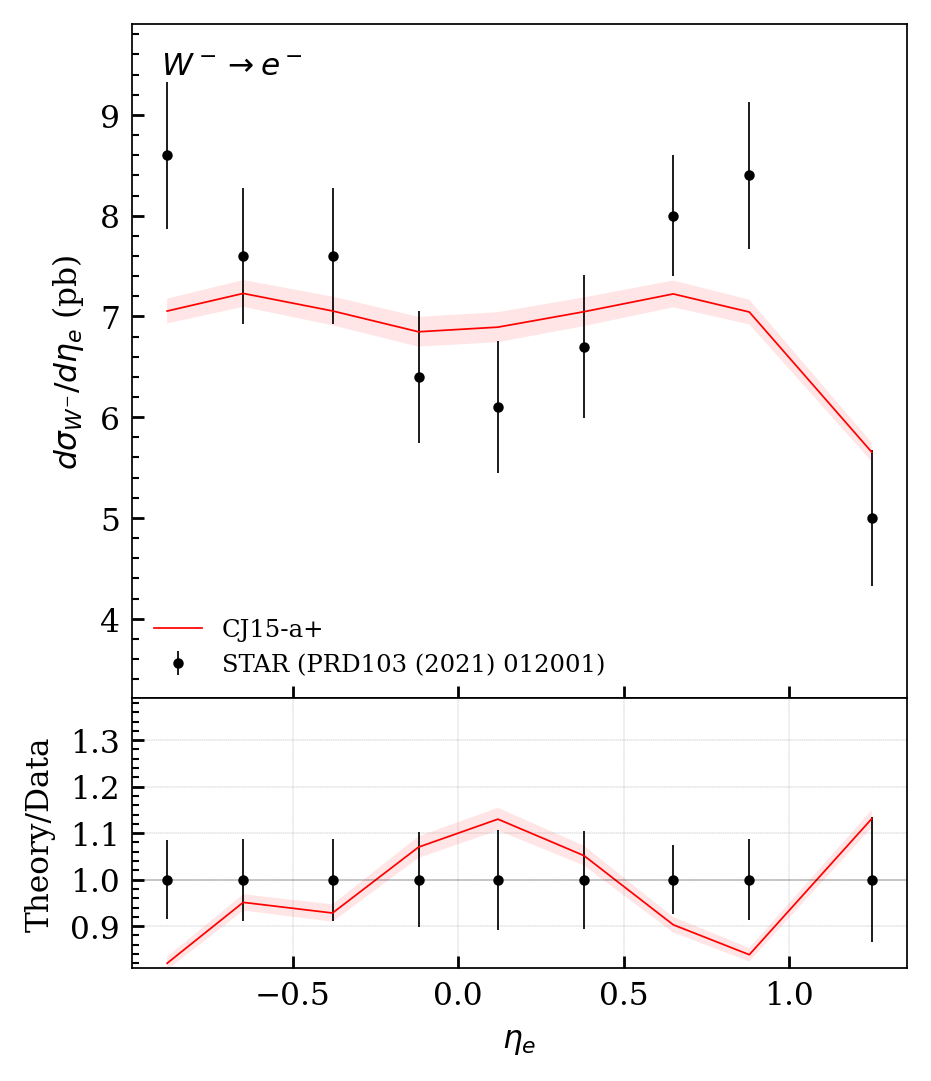}
\includegraphics[width=0.49\textwidth]{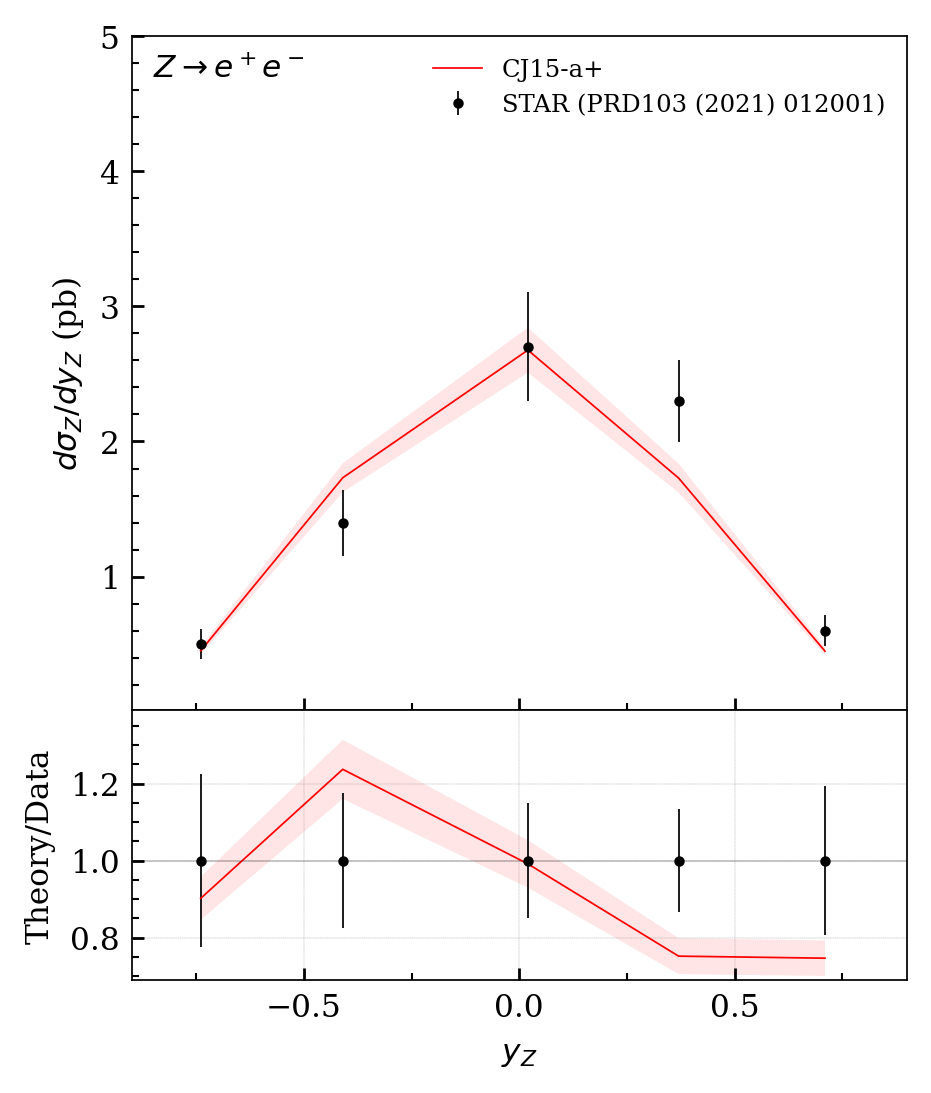}
\includegraphics[width=0.49\textwidth]{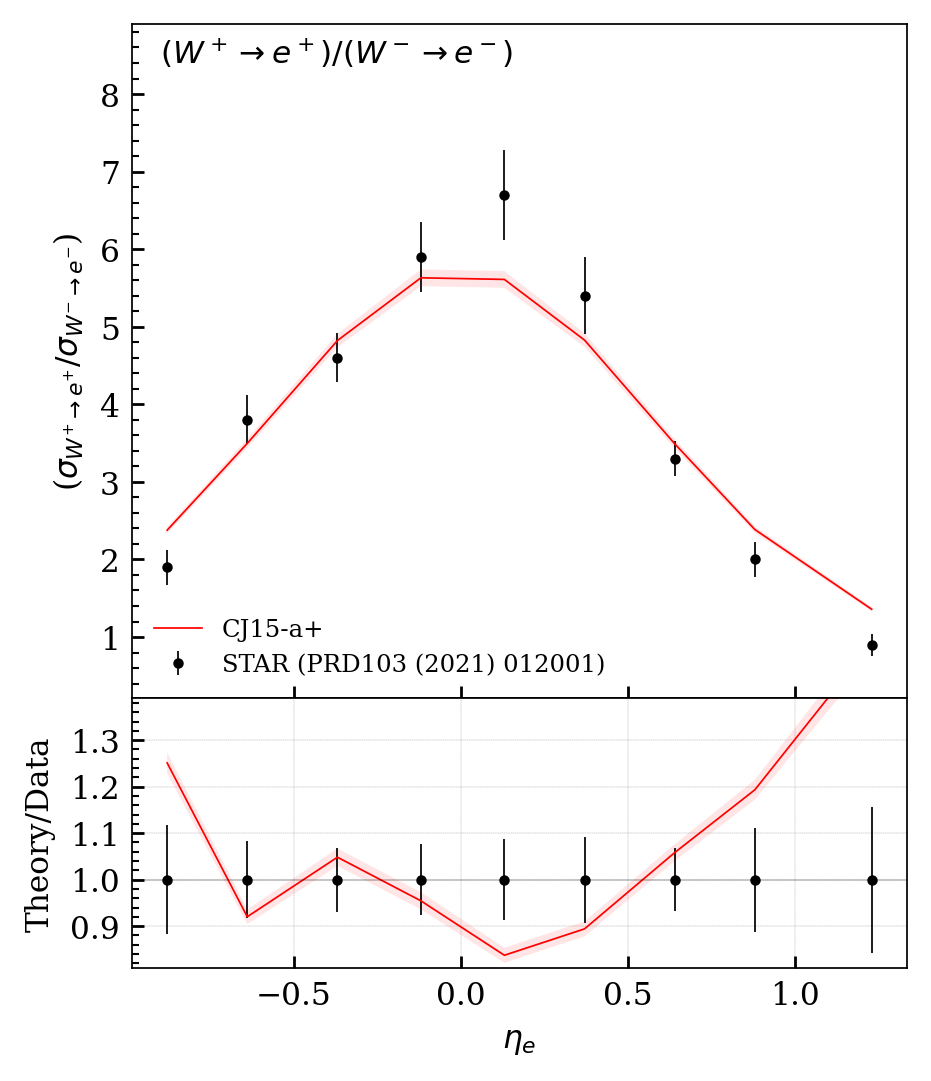}
\caption{A comparison of the \texttt{CJ15-a+} NLO calculations to the measured $W^{+}\rightarrow e^{+}$ (a), $W^{-} \rightarrow e^-$ (b), $Z \rightarrow e^+ e^-$ (c) cross sections, and $(W^{+} \rightarrow e^+)/(W^{-} \rightarrow e^-)$ charge ratio (d) by STAR. Note that only the latter was included in the PDF fit. 
}
\label{fig:obs}
\end{figure}
\noindent

Figure~\ref{fig:obs} compares the \texttt{CJ15-a+} NLO calculation of $W^{\pm}\rightarrow e^{\pm}$ and $Z/\gamma^* \rightarrow e^+ e^-$ differential cross sections and the lepton charge ratio, $\sigma_{W^+ \rightarrow e^+}/\sigma_{W^- \rightarrow e^-}$, to the STAR measurements. We used the MCFM Monte Carlo event generator \cite{MCFM} interfaced with LHAPDF \cite{lhapdf} and APPLgrid \cite{applgrid} to include NLO QCD corrections. The experimental cut of $25 < E^e_{T} < 50$ GeV for $W$ decay electrons and transverse momentum requirement of decay electrons $p^e_{T} >$ 15 GeV/c as well as the invariant mass range of 70 GeV $< M_{e^+ e^-} <$ 110 GeV for $Z$ decay electron pairs are applied respectively. As a result, a fast NLO interpolation grid was produced which was then incorporated into the CJ framework. 

The new \texttt{CJ15-a+} fit describes reasonably well the $W$ charge ratio data, which was included in the fitted data set, as well as the data for the $W^+\to e^+ $ and $Z \to e^+e^-$ differential cross sections. 
However, the measured $W^- \to e^-$ cross section shows a more pronounced oscillatory shape around midrapidity than the perturbative calculation is allowed to display by the fitted global data set. 
Furthermore, the highest pseudorapidity point in the $W^{+}$ cross section (and consequently in the lepton charge ratio) falls way below the theoretical calculation. 

\begin{figure}
    \centering
    \includegraphics[width=0.48\textwidth]{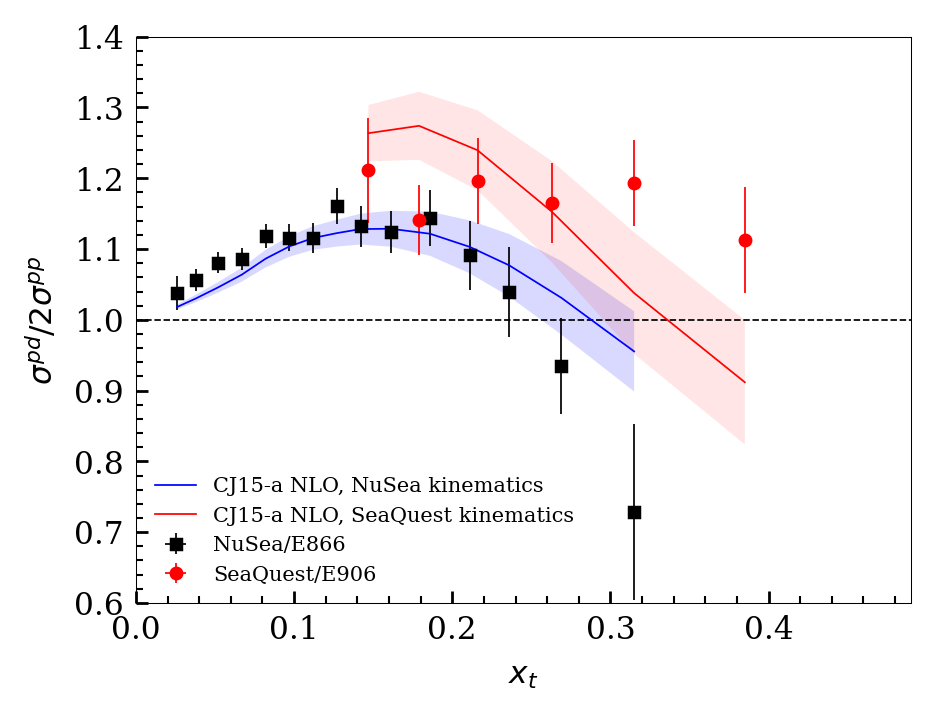}
    \includegraphics[width=0.48\textwidth]{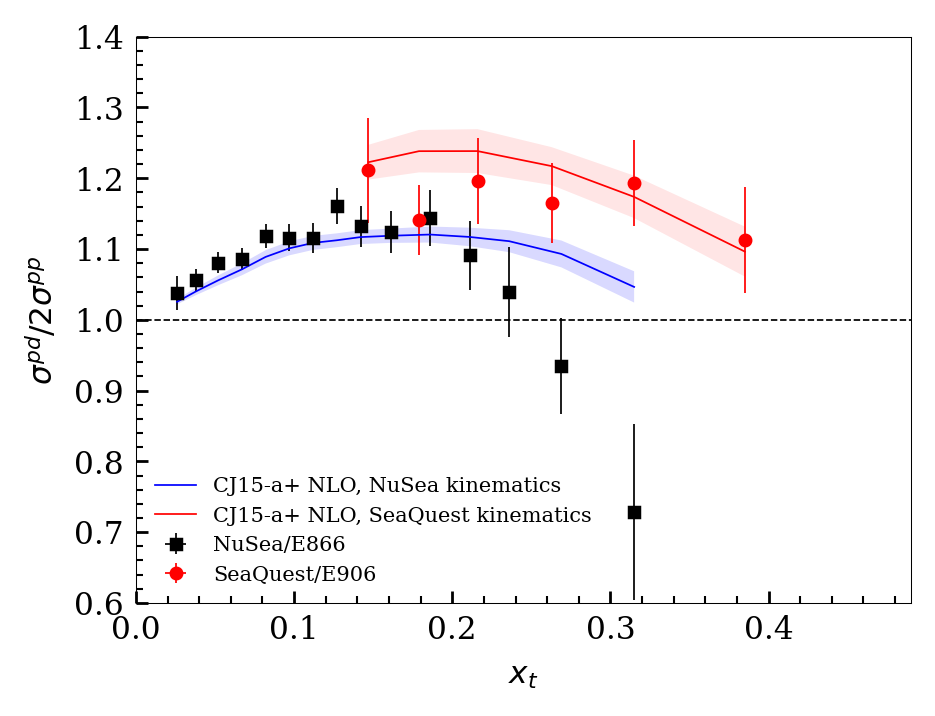}
\caption{Cross section ratio of Drell-Yan lepton pair production in $p+d$ and $p+p$ collisions by E866 and SeaQuest experiments compared with CJ15 NLO calculations before (left) and after (right) including the SeaQuest and STAR data.}
\label{fig:obs_dbub}
\end{figure}

The comparison of calculations and data for Drell-Yan lepton pair production is shown in Fig.~\ref{fig:obs_dbub} both before (left) and after (right) including the SeaQuest and STAR data into our fits. 
When only E866 data are fitted, the $\sigma^{pd}/\sigma^{pp}$ cross section ratio exhibits a steeper downturn than allowed by the SeaQuest data, see the left panel. As the right panel shows, however, including the SeaQuest data convincingly brings the ratio above 1 and substantially reduces the calculation's PDF uncertainty.

\begin{figure}[tbh]
\centering
\includegraphics[width=0.48\textwidth]{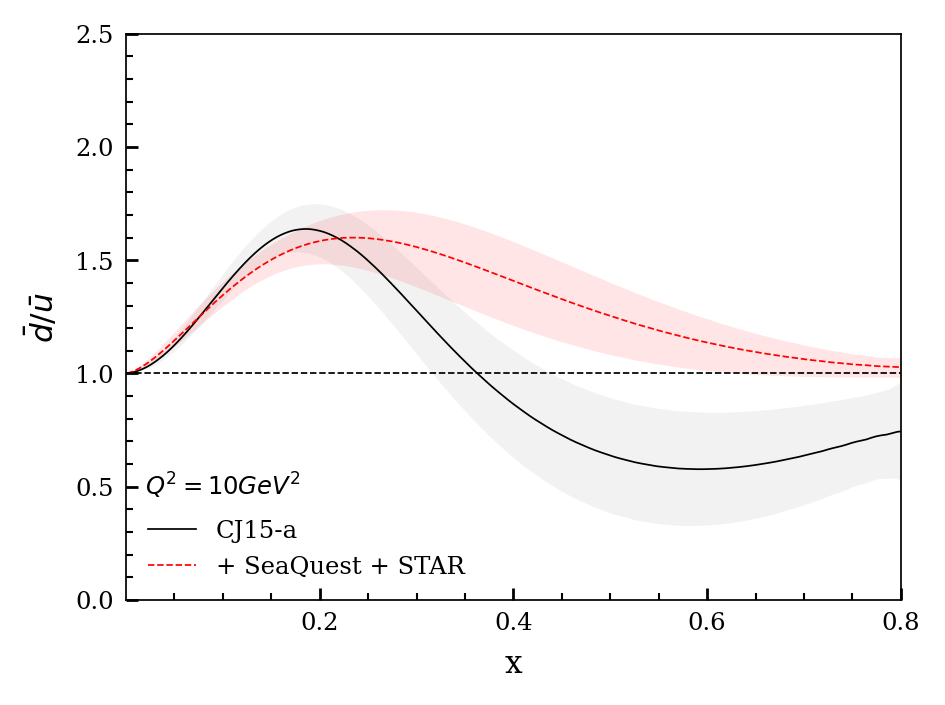}
\caption{$\bar{d}$ over $\bar{u}$ ratio at $Q^2$ = 10 GeV$^2$ extracted from the CJ15 PDF fits before (\texttt{CJ15-a}, black solid line) and after (\texttt{CJ15-a+}, red solid line) adding the STAR $W$ lepton charge ratio data in $p+p$ collisons and the SeaQuest Drell-Yan lepton pair production ratio in $p+p$ and $p+d$ collisions on top of the data sets used for the standard CJ15 PDF analysis.}
\label{fig:dbub}
\end{figure}

Finally, in Figure~\ref{fig:dbub} we plot the $\bar{d}/\bar{u}$ distribution obtained after including the STAR lepton charge ratio and the new Drell-Yan lepton pair production ratio by SeaQuest. 
Compared to the \texttt{CJ15-a} fit, which only included the E866 data, one can see that $\bar d / \bar u$ ratio remains above 1 at all values of $x$ and relaxes to 1 as the parton momentum fraction increases towards its maximum value. %
At large $x$ the antiquark ratio is pulled up by the SeaQuest data, that display a moderate $\approx 2$--$3 \sigma$  tension with respect to E866. On the contrary, STAR data are compatible with the E866 measurement around $x\approx 0.2$, where they induce a marginal suppression of the previously fitted antiquark ratio. Both experiments clearly reduce the PDF uncertainty of the $\bar d / \bar u$ ratio, with STAR data also impacting the $d/u$ ratio at the $\sim 10\%$ level (not shown here).

\section{Conclusion}

We have performed a global PDF analysis with new weak boson data by the STAR collaboration at RHIC and Drell-Yan lepton pair production ratio data by the SeaQuest experiment at Fermilab, and studied the impact of these data sets on the $\bar{d}/\bar{u}$ distribution.
This is the first time the RHIC weak boson data were included in a global unpolarized PDF analysis. Overall the sensitivity of the $W$ data is smaller than the Drell-Yan data due to the limited coverage in $x$ and relatively large experimental uncertainty, however it provides additional constraints on the light quark distributions in the intermediate $0.1 \lesssim x \lesssim 0.3$ region. With the addition of the large-$x$ SeaQuest data, the new CJ PDF fit shows that the $\bar{d}/\bar{u}$ ratio stays above unity over the whole measured region, and is compatible with $\bar d / \bar u \to 1$ as $x \to 1$.

\section*{Acknowledgements}

We would like to thank J. Bane, S. Fazio, M. Posik, and A. Tadepalli for useful discussions on their experimental measurements.


\paragraph{Funding information.}
This work was supported in part by the  U.S. Department of Energy (DOE) contract DE-AC05-06OR23177, under which Jefferson Science Associates LLC manages and operates Jefferson Lab. 
AA also acknowledges support from DOE contract DE-SC0008791.
X. Jing was partially supported by DOE Grant No. DE-SC0010129.
S. Park acknowledges support from DOE contract DE-FG02-05ER41372 and the Center for Frontiers in Nuclear Science.



\bibliography{dis_cj}

\nolinenumbers

\end{document}